\documentclass[aps,pra,twocolumn,superscriptaddress,notitlepage]{revtex4-2}

\usepackage[T1]{fontenc}
\usepackage{graphicx}
\usepackage{verbatim}
\usepackage[separate-uncertainty=true,per-mode=symbol]{siunitx}
\usepackage{amsmath, amsfonts, amssymb}
\usepackage{mathtools}
\usepackage{parskip}
\usepackage[dvipsnames]{xcolor}
\usepackage{subcaption}
\usepackage[justification=raggedright]{caption}
\usepackage{soul}

\usepackage{hyperref}
\hypersetup{%
colorlinks,
citecolor=blue,
linkcolor=blue,
urlcolor=blue
}

\usepackage[capitalize]{cleveref}
\crefname{figure}{Fig.}{Figs.}
\Crefname{figure}{Figure}{Figures}
\crefname{section}{Sec.}{Secs.}
\Crefname{section}{Section}{Sections}

\newcommand{\Rb}{$^{87}$Rb}
\newcommand{\diff}{\mathrm{d}}
\newcommand{\e}{\text{e}}
\newcommand{\im}{\text{i}}
\newcommand{\half}{\frac{1}{2}}

\renewcommand{\Re}{\operatorname{Re}}
\renewcommand{\Im}{\operatorname{Im}}

\DeclareRobustCommand{\solidLine}{\includegraphics[height=5.7pt]{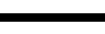}}
\DeclareRobustCommand{\dashedLine}{\includegraphics[height=5.7pt]{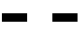}}
\DeclareRobustCommand{\finelyDashedLine}{\includegraphics[height=5.7pt]{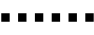}}

\DeclarePairedDelimiter{\ket}{\lvert}{\rangle}

\makeatletter
\renewcommand\tableofcontents{%
@starttoc{toc}%
}
\makeatother

\makeindex
\begin{document}

\title{Optical spectroscopy of Bose-Einstein condensates at finite temperature}

\author{R. M. F. Andersen}
\author{L. N. Stokholm}
\author{I. Zebergs}
\author{N. R. Neubert}
\author{A. S. Chatterley}
\affiliation{Center for Complex Quantum Systems, Department of Physics and Astronomy, Aarhus University, Ny Munkegade 120, DK-8000 Aarhus C, Denmark.}
\author{N. Kj{\ae}rgaard}
\affiliation{Department of Physics, Quantum Science Otago (QSO), and Dodd-Walls Centre for Photonic and Quantum Technologies, University of Otago, Dunedin, New Zealand.}
\author{J. J. Arlt}
\affiliation{Center for Complex Quantum Systems, Department of Physics and Astronomy, Aarhus University, Ny Munkegade 120, DK-8000 Aarhus C, Denmark.}

\begin{abstract}
We report on optical spectroscopic measurements of ultracold and partially condensed $^{87}$Rb gases, which show distinct spectral features due to the thermal and the Bose condensed components in the frequency domain. These features are detected in-situ by using a dark-field configuration with single-photon sensitivity and a frequency-agile laser system. To interpret the observed spectra, we develop a model for light propagation through an inhomogeneous atomic cloud. This model enables the extraction of temperature and atom number, which we benchmark against conventional time-of-flight absorption imaging. The spectroscopically obtained cloud parameters show enhanced sensitivity to small thermal fractions in nearly pure condensates. We further compare the spectroscopy in dark-field and bright-field configurations, demonstrating the superior performance of the former. Our results reveal previously unexplored spectral structure in optically dense ultracold gases and establish spectroscopy as a tool for characterizing ultracold systems at very low temperatures.
\end{abstract}
\maketitle

\section{Introduction}
\label{sec. Introduction}

The realization of Bose–Einstein condensation (BEC) in dilute atomic gases has had a significant impact on modern physics, providing an unprecedented platform for exploring quantum phenomena on macroscopic scales~\cite{Pethick2008,pitaevskii_bose-einstein_2003}. The high degree of experimental control available in ultracold atomic systems, including external potentials, dimensionality, and interactions, has enabled their application in quantum simulation, fundamental physics, and in particular for the investigation of many-body physics~\cite{Bloch2008,pitaevskii_bose-einstein_2016}. Thus, the continued development of accurate probing techniques for BECs is essential for understanding these strongly correlated quantum systems.

The detection of ultracold atomic clouds is typically achieved by optical methods, based on the interaction of the constituent atoms with light. These methods can broadly be divided into resonant and dispersive techniques. These techniques have been reviewed in detail recently~\cite{wolswijk2025,Vibel2024}.

Resonant imaging techniques rely on detecting photons resonantly scattered by the atomic sample, with absorption and fluorescence imaging being the most widely used approaches. In both cases, the atom number is inferred from the number of photons scattered. Fluorescence imaging, which directly collects this scattered light, has seen widespread use across atomic physics experiments. Absorption imaging, in particular, has become a cornerstone of ultracold atomic physics, including investigations of Bose–Einstein condensation~\cite{Ketterle1999, Reinaudi2007_alpha}. Notably, it can also be implemented in a dark-field configuration, where spatial filtering in the Fourier plane suppresses the unscattered probe beam, enhancing sensitivity to weak signals~\cite{Pappa2011,Reinhard2014}.

Dispersive imaging techniques, in contrast, measure the phase shift imprinted on off-resonant probe light by the spatially varying refractive index of the atomic cloud~\cite{KetterleDF,HuletDF}. Since they do not rely on photon absorption, these methods can be made minimally destructive, enabling repeated probing of the same ultracold sample. The resulting signal depends sensitively on the density-dependent refractive index of the gas~\cite{GajdaczNonDestructive}. In practice, such techniques are typically operated at a fixed, non-zero detuning, necessitating a degree of spectroscopic understanding of the system—even when a complete spectral characterization is not explicitly performed.

At low temperatures and large condensate fractions, it becomes increasingly difficult to extract the thermal atom number and temperature using imaging techniques, since the small thermal component is typically obscured by the BEC. Nonetheless, absorption imaging can be used to characterize very cold finite temperature BECs~\cite{Leanhardt2003}, which can be exploited in low-gravity environments~\cite{temperatureLowGravity}. Alternative techniques include using embedded impurities such as atoms~\cite{temperatureImpurityThermometry1} or magnons~\cite{temperatureMagnonThermometry}, and Bragg scattering~\cite{temperatureBragg2004}. 

Although optical spectroscopy is the standard tool for probing atomic systems, the complex spectral signatures that arise in partially condensed ultracold gases have received relatively limited attention. The most important previous application was the spectroscopic detection of the condensation of atomic hydrogen~\cite{spectrumPaper4_HydrogenBEC}. Moreover, optical spectroscopy has been applied in cold atomic gases to investigate dipolar interactions~\cite{spectrumPaper5_SrAtoms,spectrumPaper1_Browaeys2016, spectrumPaper2_Browaeys2014} or electromagnetically induced transparency~\cite{otherSpectra_Hau, otherSpectra_Lindsay2019, spectrumPaper3_EITThermometry}.

This paper examines optical-frequency spectroscopy of partially condensed ultracold atomic clouds at the single-photon level. We observe rich spectral features that arise from the large in-situ optical densities of the samples. The spectra are investigated using a dark-field (DF) configuration, where the unscattered light is suppressed, as well as a bright-field (BF) setup, where the transmitted light is directly measured. The DF method is sensitive to both absorption and phase-shift, resulting in enhanced sensitivity to the system parameters. The spectroscopic measurements are compared with a theoretical model, which is also used to extract the thermal atom number, condensed atom number, and temperature. These results are benchmarked against absorption imaging to validate the method. An advantage of using spectroscopy is that it can be measured in-situ, eliminating the need to model expansion dynamics, which is especially difficult in the high condensate fraction regime~\cite{expansionPaper}. The non-destructive nature of the spectroscopic method also allows for repeated measurements of the same cloud. Furthermore, the spatial resolution requirements for obtaining spectroscopic information are less stringent than for imaging methods.

This manuscript is organized as follows. \Cref{sec. Spectroscopic detection} introduces the experimental techniques and theoretical framework necessary to measure the spectra and extract cloud parameters. It also discusses the novel spectral features caused by high in-situ optical densities of ultracold clouds. \Cref{sec. Spectroscopic fitting} compares the extracted cloud parameters from fitted spectra to the established TOF imaging method and illustrates the advantages of applying the spectroscopic method for very cold partially condensed clouds. \Cref{sec. Conclusion and outlook} summarizes the results and discusses its applicability in future experiments.

\section{Spectroscopy of partially condensed atomic clouds}
\label{sec. Spectroscopic detection}

\subsection{Experimental setup}
\label{subsec. Optics setup}

The experiments are performed on clouds of \Rb{} atoms prepared in the $\ket{F=2, m_{\text{F}}=2}$ state and confined in a crossed optical dipole trap (cODT) at a wavelength of $1064$~nm with approximately $\SI{80}{\micro\meter}$ beam waists. Using standard evaporative cooling techniques, BECs containing up to $6.5\cdot10^5$ atoms can be produced. The condensate fraction is controlled by adjusting the final dipole trap power. A detailed description of the preparation sequence can be found in~\cite{MalthePhD, LaurtisPhD}.

\begin{figure}[t]
\centering
\includegraphics[width=\linewidth]{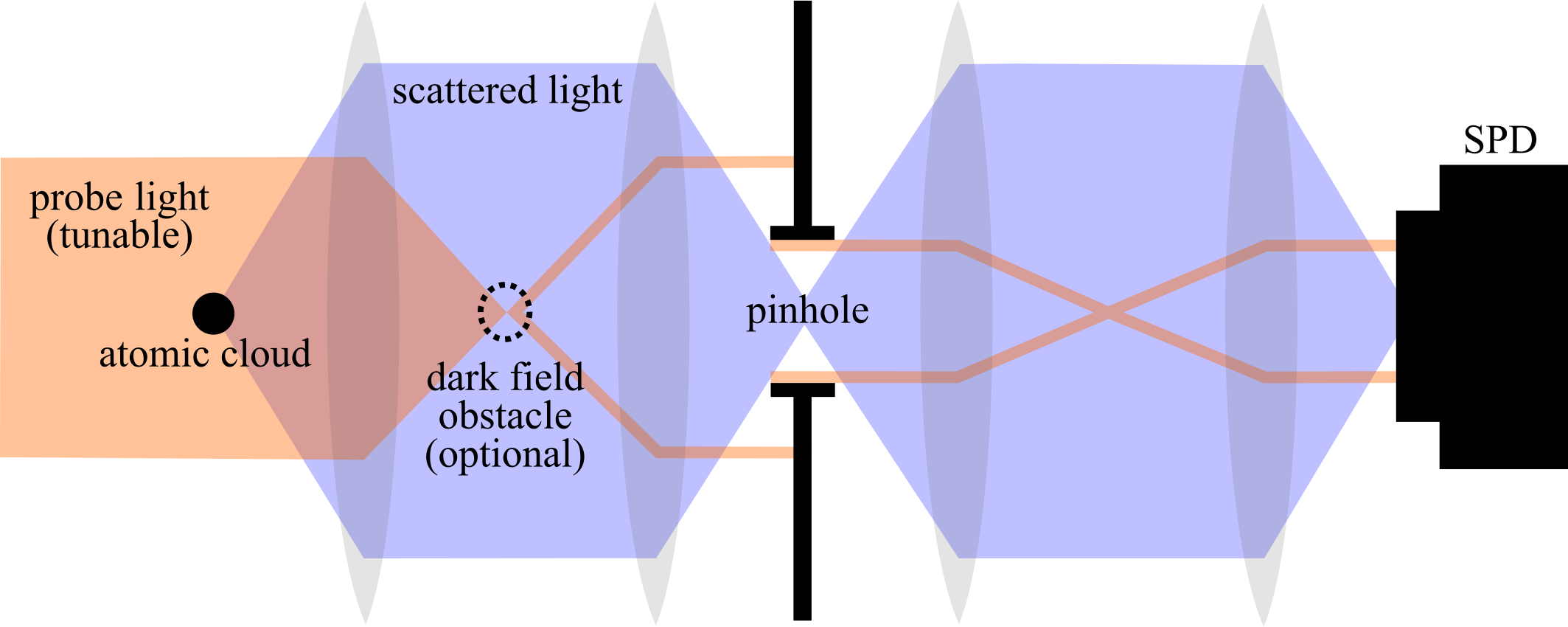}
\caption{Schematic of the detection setup. Collimated probe light (orange) is sent through a cloud of trapped atoms, and some of the light is scattered (blue). At the focus plane of the probe beam, an optional small dark-field obstacle can be placed to block only the unscattered light. Additional filtering is achieved by a pinhole at an intermediate image plane. The image is either relayed onto a camera or focused onto a single-photon detector (SPD).}
\label{fig. imagingOptics}
\end{figure}

To perform spectroscopy, $\sigma^+$-polarized probe light drives the cycling transition between the $\ket{2,2}$ and the $\ket{3,3}$ excited state. The probe laser is beat-locked to a reference laser using an optical phase-lock loop (OPLL)~\cite{JurgenAppel}, enabling agile frequency detuning on a \SI{9}{\giga\hertz} frequency range with a maximal sweep rate of \SI{90}{\mega\hertz\per\milli\second}. A peak probe field intensity of $\approx 10^{-6}I_{\mathrm{sat}}$, is used, where $I_{\mathrm{sat}}$ is the saturation intensity. Thus, on average, at most one photon is present in the cloud at any moment. The spatial mode of the probe field is much larger than the cloud. 
The transmitted probe light is relayed through an imaging system onto a single-photon detector (SPD) connected to a homemade time-to-digital converter capable of recording millions of events over arbitrary periods of time. The setup is shown in~\cref{fig. imagingOptics}. Photon arrival times recorded by the SPD are binned during post-processing. The bin size is adjusted across the frequency scan to appropriately resolve the spectral structures while minimizing the shot noise. Since the probe frequency is swept in time via a frequency generator and the OPLL, the time axis is directly mapped to frequency.

The imaging system can operate in either BF or DF configuration. In the DF configuration, a metal wire of $\SI{200}{\micro\meter}$ diameter is placed at the Fourier plane to block unscattered light~\cite{Pappa2011}. The obstacle is mounted on a translational stage for easy insertion and removal. In addition, a pinhole with $\SI{150}{\micro\meter}$ diameter is positioned in an intermediate image plane to spatially filter the signal. This pinhole selects an effective area around the atoms that contributes to the spectrum. In the BF configuration, this pinhole is mandatory to avoid saturation of the SPD, and in the DF configuration, it suppresses residual transmitted light further.

The numerical aperture (NA) of the optical system is $0.18$, corresponding to a diffraction-limited resolution of $\SI{2.6}{\micro\meter}$ according to Rayleigh’s criterion. This resolution is insufficient to resolve the in-situ atomic density distribution. This effect is accounted for in the model described in \cref{subsec. Calibration}.

In addition, the experiment is equipped with a CCD camera in an orthogonal imaging axis, which facilitates standard time-of-flight (TOF) absorption imaging~\cite{Ketterle1999, Reinaudi2007_alpha} for benchmarking cloud parameters.

\subsection{Model of ultracold cloud spectra}
\label{subsec. Spectral theory}
In the following, a model for ultracold cloud spectra in both BF and DF configurations is developed. It is used to discuss the expected spectral features at high optical densities. 

\begin{figure*}
\centering
\begin{subfigure}[]{0.35\linewidth}
\includegraphics[width = 1\linewidth]{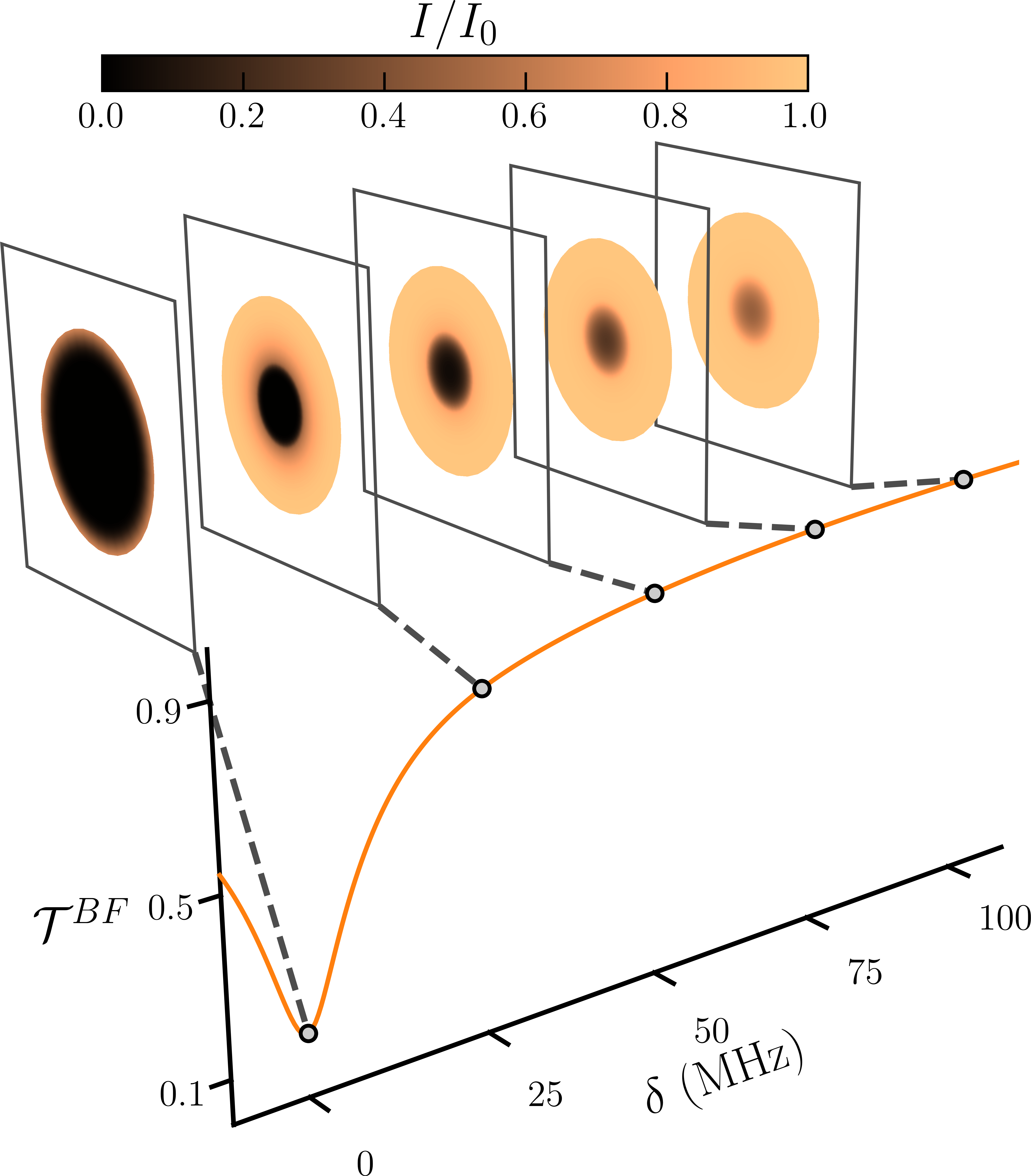}
\end{subfigure}
\hspace{{5em}}
\begin{subfigure}[]{0.35\linewidth}
\includegraphics[width = 1\linewidth]{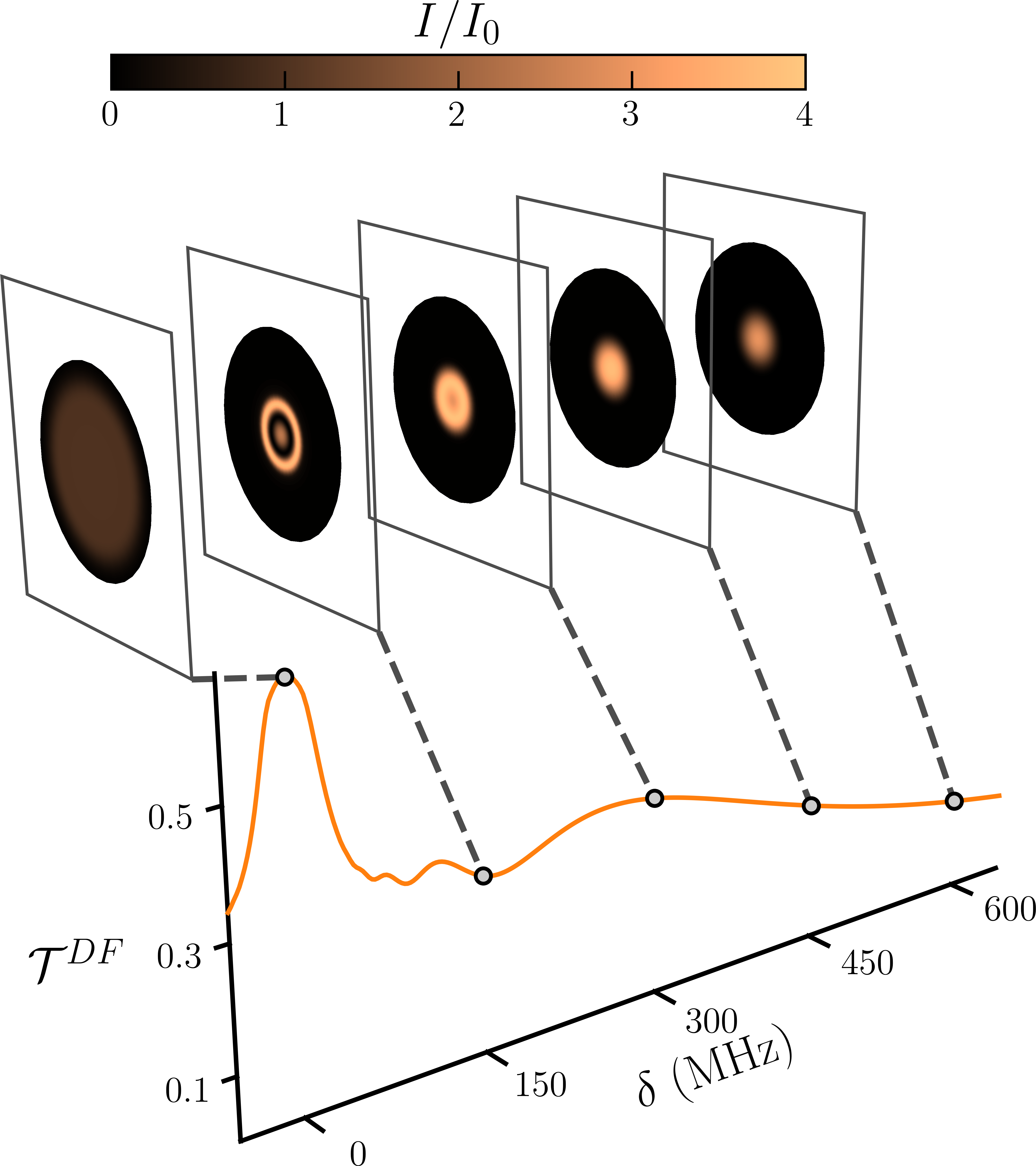}
\end{subfigure}
\caption{Simulated BF (left) and DF (right) spectra of a partially condensed cloud at a temperature of $\SI{130}{\nano\kelvin}$ containing $1.70\cdot10^5$ condensed atoms and $5\cdot10^4$ thermal atoms, in a symmetrical harmonic trap with trap frequencies $2\pi \cdot \SI{100}{\hertz}$. Insets show the intensity profiles of the transmitted light and the scattered light (see~\cref{fig. imagingOptics}) immediately after the cloud at various detunings (see text). The circular area corresponds to the pinhole, with a radius chosen for illustrative purposes. Note, that the color scale varies between the two plots.
}
\label{fig. spectrum visualisation}
\end{figure*}

Intensity profiles immediately after a partially condensed cloud are illustrated in~\cref{fig. spectrum visualisation} at various detunings of the probe light $\delta$. The field after the atoms consists of the sum of the unperturbed field and a scattered light field. In the BF configuration, the shown intensity profile corresponds to the total light field that appears after passing through the cloud. In the DF configuration, we only show intensity corresponding to the scattered light field, since this is measured in the DF setup (see~\cref{fig. imagingOptics}).

The incoming probe light propagates along the $+z$ direction, has an intensity of $I_0$, and is assumed to be constant over the pinhole region, which has an effective area $A$ accounting for magnification. The transmitted light power recorded by the SPD in BF is normalized to the incoming probe power $I_0A$ and denoted by $\mathcal{T}^{BF}$. Thus,
\begin{align}\label{eq. normalized T BF}
\mathcal{T}^{BF}=\frac{1}{A}\int_{\text{pinhole}} t(x,y)^2\ \diff x \diff y,
\end{align}
where $t(x,y)$ is the spatially dependent electric field transmission coefficient through the cloud. This can be calculated as
\begin{align}
t(x,y) = \exp \left(-\int_{-\infty}^\infty\Im\left[n(\mathbf{x})\right]k\diff z\right),
\label{eq. E-field transmission BF}
\end{align}
where~\cite{jackson1999classical, StratenDF}
\begin{align}
n(\mathbf{x})=\sqrt{1+\frac{\chi(\mathbf{x})}{1-\frac{1}{3}\chi(\mathbf{x})}},
\end{align}
is the complex index of refraction according to the Lorentz-Lorenz model, and the electric susceptibility $\chi(\mathbf{x})$ is given by~\cite{Boyd-NonlinearOptics}
\begin{align}
\chi(\mathbf{x})=-\frac{\sigma_0\rho(\mathbf{x})}{k}\frac{1}{2\Tilde{\delta}+\im},
\end{align}
where $\sigma_0$ is the resonant photon absorption cross section~(taking into account the relevant Clebsch-Gordan coefficients), $\rho$ is the number density of atoms, $\Tilde{\delta}=(\omega-\omega_0)/\Gamma$ is the normalized detuning with $\omega_0$ being the resonance frequency and $\Gamma$ is the natural linewidth FWHM of the transition. The probe light intensity is assumed to be much lower than the saturation intensity. Effects due to bosonic statistics~\cite{Konstantinou_2026, morice1995refractive, Bons_2016, Lu2023Bosonic} are not included in this model, and neither are dipole-dipole interactions~\cite{spectrumPaper2_Browaeys2014, spectrumPaper1_Browaeys2016, DipoleDipoleCloudShapeChange} or the incoherent emission signal from the atoms~\cite{Pappa2011}.

The number density is given by $\rho = \rho_0 + \rho_{th}$ where the BEC density $\rho_0$ is~\cite{Pethick2008}
\begin{align}\label{eq. BEC density}
\rho_0(\mathbf{x}) = \frac{15}{8\pi}\frac{N_0}{R_1R_2R_3}\cdot \max\left( 0, 1 - \sum_{i=1}^{3} \frac{x_i^2}{R_i^2} \right),
\end{align}
where $x_i$ refer to the three directions with corresponding Thomas-Fermi radius $R_i = \sqrt{2\mu/(m\omega_i^2)}$, and the chemical potential $\mu$ related to the condensed atom number in the Thomas-Fermi approximation by $\mu = \half \left(15N_0a/a_{\text{ho}}\right)^{2/5}\hbar \Bar{\omega}$. Here, $a$ is the $s$-wave scattering length, $a_{\text{ho}}=\sqrt{\hbar/(m\Bar{\omega})}$ is a characteristic length in the harmonic oscillator, and $\Bar{\omega}=(\omega_1\omega_2\omega_3)^{1/3}$ is the geometric mean of the trap frequencies. The thermal atom density $\rho_{\mathrm{th}}$ is given by the semi-ideal model~\cite{Naraschewsk1998}
\begin{align}\label{eq. time dependent thermal cloud}
\rho_{\mathrm{th}}(\mathbf{x})=\lambda_T^{-3}g_{\frac{3}{2}}\left(\e^{\big|V(\mathbf{x})-\mu\big|/k_BT}\right),
\end{align}
where $\lambda_T=\big[2\pi\hbar^2/(mk_BT)\big]^{1/2}$ is the thermal de-Broglie wavelength, $V(x_i)$ the trapping potential, and where $g_\gamma(x)$ is the polylogarithm function.

The resulting BF spectrum, shown in the left panel of \cref{fig. spectrum visualisation}, consists of contributions from both the condensed and thermal parts of the cloud, characterized by their optical densities ($od$) and sizes. The condensate has high $od$, and hence primarily light at large detuning penetrates it, resulting in a broad spectroscopic feature. However, the condensate is typically small, and consequently, the total absorption is limited. The thermal component has larger tails, but a lower $od$, and thus contributes a narrow spectroscopic feature with greater resonant absorption. The spectral features from both components blend together and are typically hard to distinguish qualitatively in the BF spectrum.

The right panel of \cref{fig. spectrum visualisation} illustrates the scenario in the DF configuration. A series of scattered light field intensity profiles immediately after the atomic cloud is shown at various detunings, together with the spectroscopic signal measured on the SPD. The DF signal, normalized to the incoming probe power $I_0A$, is given by~\cite{Ketterle1999}
\begin{align}
\label{eq. normalized T DF}
\mathcal{T}^{DF}=\frac{1}{A}\int_{\mathclap{\raisebox{-2ex}{\scriptsize pinhole}}}\   1+t(x,y)^2-2t(x,y)\cos\left[\phi(x,y)\right] \ \diff x\diff y,
\end{align}
where $t(x,y)$ is the electric field transmission coefficient from \cref{eq. E-field transmission BF}, and $\phi(x,y)$ is the spatially dependent phase-shift, relative to vacuum, given by
\begin{align}\label{eq. phase shift DF}
\phi(x,y)=k\int_{-\infty}^\infty  \Re\big[n(\mathbf{x})\big]-1 \ \diff z.
\end{align}
Because DF is sensitive to phase shifts, the spectra become much broader and contain more complex structures. The spectrum is separated into a strong, narrow peak caused by the thermal component and a broad structure caused by the BEC for the same reasons as were discussed in the BF case. Since the peak phase-shift is typically several $\pi$, the spectrum is no longer monotonic. A noticeable feature is the large shoulder at the intermediate detuning around $\SI{150}{MHz}$. This occurs when a ring of phase-shift equal to $\pi$ appears such that the integrated signal is maximized, while the peak phase-shift at the center is not enough to reach $2\pi$. Thus, the spectroscopic features caused by the thermal and condensed components are qualitatively different, as shown by the spectrum in \cref{fig. spectrum visualisation}. The bimodal structure in the spectrum can be compared to the spatial bimodal density profile typically observed in TOF images, however, with reversed roles.

Thus, both BF and DF spectra contain information about the spatial distribution of the clouds, including the thermal atom numbers, BEC atom numbers, and the temperature. This probing method can be used on a wide range of clouds since the optical densities probed span from the resonant value to almost zero over the entire spectrum. \cref{sec. Spectroscopic fitting} investigates how well these parameters can be extracted.

The results given by~\cref{eq. normalized T BF} and~\cref{eq. normalized T DF} can be applied in traps with non-harmonic geometries by calculating the appropriate transmission and phase-shift. In BF, it is possible to include an inhomogeneous probe beam intensity across the atomic cloud. However, for DF it is necessary that the probe beam is much larger than the cloud, such that the unperturbed light field can be filtered effectively from the scattered light field in the Fourier plane. Finally, the model essentially uses the thin-lens approximation for the atomic cloud to calculate the light propagation, and a full light-propagation simulation may lead to more accurate predictions~\cite{Deb_2020}.

\begin{figure}[t]
\centering
\includegraphics[width=\linewidth]{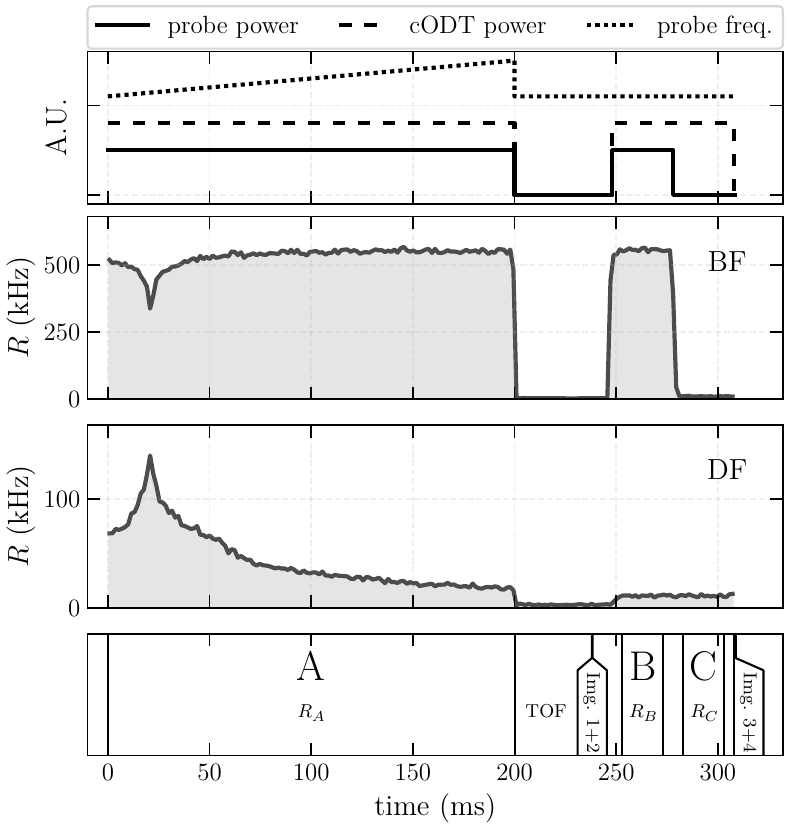}
\caption{Measurement sequence of spectra and the following absorption images after TOF. (a) Illustrative traces of the probe power \mbox{(\solidLine)}, cODT power \mbox{(\dashedLine)}, and probe frequency \mbox{(\finelyDashedLine)}. (b) Typical count rate observed on the SPD in BF configuration. (c) Count rate on SPD observed in DF configuration. (d) Time intervals in the experimental sequence. The three intervals labelled A, B, and C, are used to calculate $\mathcal{T}^{BF/DF}$ (see text). }
\label{fig. BOTH data acquisition}
\end{figure}

\subsection{Experimental measurement sequence}
\label{subsec. Measurement sequence}

The experimental sequence used to acquire spectra is illustrated in~\cref{fig. BOTH data acquisition}. \cref{fig. BOTH data acquisition}(a) shows the cODT power, the probe power, and the probe frequency during a measurement sequence. Typical BF and DF spectra are shown in~\cref{fig. BOTH data acquisition}(b) and ~\cref{fig. BOTH data acquisition}(c) respectively. The signal $R$ is derived from the individual photon arrival times recorded by the SPD. The photon arrival times are binned into intervals, and the recorded photon number in each bin is divided by the bin duration, giving the rate $R$ of detected photons. Measurements during three time intervals, indicated with A, B, and C in~\cref{fig. BOTH data acquisition}(d) are used to obtain the normalized transmission spectra. 

In interval A, the frequency of the probe light is swept with constant power in the presence of the cODT to record $R_A$, yielding the spectra. The signal without atoms, but with probe and trapping light on, is denoted $R_B$. This signal is used to calculate the probe intensity when the setup is in BF configuration and the background signal when the setup is in DF configuration. Finally, the signal without the probe light, but with the trapping light on, is denoted $R_C$ and is used in the BF configuration to estimate the background signal. Superscripts $BF$ or $DF$ are used to indicate the configuration of the detection system while acquiring the data, and a bar-symbol ($\overline{\phantom{ }\cdot\phantom{ }}$) indicates the mean signal in a given interval.

The normalized transmission signal for BF is then given by
\begin{align}\label{eq. BF raw data to normalized T}
\mathcal{T}^{BF}=\frac{R_A^{BF}-\overline{R}_C^{BF}}{\overline{R}_B^{BF}-\overline{R}_C^{BF}}.
\end{align}
The DF measurement is similar, but requires a separate measurement of the probe light intensity due to the presence of the DF obstacle. The normalized transmission signal is given by
\begin{align}\label{eq. DF raw data to normalized T}
\mathcal{T}^{DF}=\frac{R_A^{DF}-\overline{R}_B^{DF}}{\overline{R}_B^{BF}-\overline{R}_C^{BF}}.
\end{align}
\Cref{fig. BOTH data acquisition} shows that the majority of the background noise in DF configuration, during interval B and C, is due to leaking photons from the cODT and not from the probe light. All signals, together with the images after TOF, are measured on the same cloud, which requires the probe light intensity to be as low as possible to cause only minimal destruction.

For all measurements discussed here, the light probes the cycling transition $\ket{2,2}\leftrightarrow\ket{3,3}$, but probing other transitions is equally possible. The frequency sweeps are chosen to span from $-\SI{160}{MHz}$ to $\SI{1440}{MHz}$ with respect to the resonance frequency. Thus, enough of the spectrum is covered to extract the necessary information without approaching the $\ket{2,2}\leftrightarrow\ket{3,2}$ transition, which leads to a spurious signal due to small polarization impurities. In the following, the normalized transmissions $\mathcal{T}^{BF}$ and $\mathcal{T}^{DF}$ allow for a comparison between the experimental results and the theoretical spectra. Further details are av

\subsection{Time-of-flight image analysis}
\label{subsec. Time-of-flight image analysis}
Absorption imaging after TOF is only outlined briefly here. It provides spatial information of the light intensity after passing through the atomic cloud. The optical density $od=\sigma_0\Tilde{\rho}$, is obtained directly from absorption images using the method described in~\cite{Reinaudi2007_alpha}, where $\Tilde{\rho}$ is the atom number column density. In~\cref{fig. BOTH data acquisition}, the timing for the acquisition of these images is indicated.

The cloud parameters $\mu$ and $T$ are extracted from the TOF images using a density distribution according to the semi-ideal model~\cite{Naraschewsk1998}. The implemented procedure automatically accounts for temperatures above and below $T_c$.

The fit function for the thermal column density distribution is given by
\begin{align}
\label{eq. thermal column density}
\Tilde{\rho}_{\mathrm{th}}= 
\begin{cases}
c_+ g_2\Bigg(\exp\Big[{-\sum_{i=1}^2\frac{(x_i-x_{i,0})^2}{2\beta_i^2w_i^2}}\Big]\Bigg),&\mu>0 \\
&\\
c_- g_2\Bigg(\exp\Big[{\frac{\mu}{k_BT}-\sum_{i=1}^2\frac{(x_i-x_{i,0})^2}{2\beta_i^2w_i^2}}\Big]\Bigg),&\mu\le0,
\end{cases}
\end{align}
where \mbox{$w_i=1/\omega_i\cdot\sqrt{k_BT/m}$}, \mbox{$\beta_i=\sqrt{1+\omega_i^2t^2}$}, $t$ is the TOF, $x_{i,0}$ are the cloud center coordinates, \mbox{$c_+=N_{th}(\mu, T)/[2\pi w_x\beta_x w_y\beta_y g_3(1)]$}, and \mbox{$c_-=\sqrt{2\pi}w_z/(\lambda_T^3\beta_x\beta_y)$}. 

Importantly, the fit function naturally distinguishes between purely thermal ensembles with $\mu\le0$ and bimodal clouds with $\mu > 0$. For $\mu\le0$, the fit function is only given by~\cref{eq. thermal column density}, and for $\mu > 0$, a BEC is present, which must also be included in the total column density.

The fit function for the BEC column density after TOF is given by
\begin{align}\label{eq. BEC column density}
 \Tilde{\rho}_0 = \frac{5}{2\pi}\frac{N_0(\mu)}{R_1R_2}\cdot \max\left[0,  1 - \sum_{i=1}^2\frac{(x_i-x_{i,0})^2}{R_i^2} \right]^{3/2},
\end{align}

where the condensed atom number is $N_0(\mu) = \frac{1}{15} \frac{a_\text{ho}}{a} \left( \frac{\mu}{\hbar \Bar{\omega}} \right) ^\frac{5}{2}$, and the BEC radii after TOF are $R_i$. The cloud center positions, $x_{i,0}$, are the same as the ones used for the thermal column density. 
The thermal atom number $N_{\mathrm{th}}(\mu,T)$ is consistently calculated by numerically integrating $\rho_{\mathrm{th}}$ from~\cref{eq. time dependent thermal cloud}~\cite{Naraschewsk1998}.

The fit is performed using the total column density $\Tilde{\rho} = \Tilde{\rho}_0 + \Tilde{\rho}_{\mathrm{th}}$ with six free parameters $\mu$, $T$, $R_i$ and $x_{i,0}$. The thermal and condensed atom numbers are then calculated from $\mu$ and $T$.

\subsection{Spectral model calibration}
\label{subsec. Calibration}

Two parameters are calibrated to take the experimental imperfections into account. Firstly, the resonant absorption cross-section $\sigma=\sigma_0/\alpha$ is scaled by a parameter $\alpha$ relative to the theoretical value $\sigma_0$. This is analogous to the usual calibration necessary for absorption imaging~\cite{Reinaudi2007_alpha}, and accounts for numerous imperfections in the light, such as polarization contamination, causing a reduction in the interaction strength. Secondly, the finite numerical aperture must be taken into account.

The aperture is incorporated by propagating the scattered electric field $E_{sc}$ (the blue field in~\cref{fig. imagingOptics}) from the atoms to the first Fourier plane according to \cite{Goodman1998}
\begin{equation}\label{eq: E in fourier plane}
E_\mathcal{F}(x,y) =   \frac{1}{\lambda f}\int E_{sc}(\tilde{x}, \tilde{y}) e^{-i \frac{2 \pi (x \tilde{x} + y \tilde{y})}{\lambda f}} \diff \tilde{x} \diff \tilde{y}
\end{equation}
where $f$ is the focal length of the first lens in the setup, $\lambda$ is the wavelength of the light, and $E_{sc}=E_0(1-\e^{-t+\im\phi})$ is the scattered electric field just after the atoms with incoming electric field strength $E_0$, $t$ given by~\cref{eq. E-field transmission BF} and $\phi$ given by~\cref{eq. phase shift DF}. Since $t$ and $\phi$ depend on light detuning, the limited numerical aperture alters the shape of the spectra. Assuming that $E_\mathcal{F}$ is collimated between the first and the second lens, any apertures in this region can be taken into account by considering their effect on $E_\mathcal{F}$, even if those apertures are not positioned in the Fourier plane.

The fraction of power $\mathcal{P}^{DF}$ reaching the detector in DF configuration is calculated by integrating the scattered light over a circular area of radius $R_\text{NA}$ in the Fourier plane while removing the stripe blocked by the DF obstacle of width $d_\text{DF}$
\begin{equation}
\label{eq: E in fourier plane}
\begin{aligned}
\mathcal{P}^{DF} &= \frac{ \int_\mathcal{A} \left| E_\mathcal{F}(x,y) \right|^2 \diff x \diff y} {\int_{\text{pinhole}} \left| E_{\mathrm{sc}}(x,y) \right|^2 \diff x \diff y },\\
\mathcal{A} &=  \left\{ (x,y)\in \mathbb{R}^2 \Big| x^2 + y^2 \leq R_\text{NA}^2, |x| > \frac{d_\text{DF}}{2} \right\} .
\end{aligned}
\end{equation}
Here, $R_\text{NA}$ serves as a calibration parameter.

In BF, the unscattered electric field is assumed not to be obstructed by the aperture, while the power lost from the scattered field is the same. Thus, the aperture transmission $\mathcal{P}^{BF}$ can be calculated from $\mathcal{P}^{DF}$ evaluated at $d_\text{DF}=0$ as follows
\begin{equation}\label{eq. aperture transmission function BF}
\mathcal{P}^{BF} = 1 - (1-\mathcal{P}^{DF})\frac{\mathcal{T}^{DF}}{\mathcal{T}^{BF}}.
\end{equation}

To calibrate $R_{\text{NA}}$ and $\alpha$, two DF datasets representing different cloud parameters are fitted with varying values of the calibration parameters, and the extracted cloud parameters are compared to the values from TOF imaging. The calibration parameters are optimized for the best agreement between the two techniques, resulting in the optimal values $R_{\text{NA}} = \SI{3.6}{mm}$ and $\alpha = 1.27$. The value of $R_{\text{NA}}$ is smaller than any physical aperture in the setup. It functions as an effective parameter encoding all losses of the scattered electric field.

\begin{figure}[t]
\centering
\includegraphics[width=\linewidth]{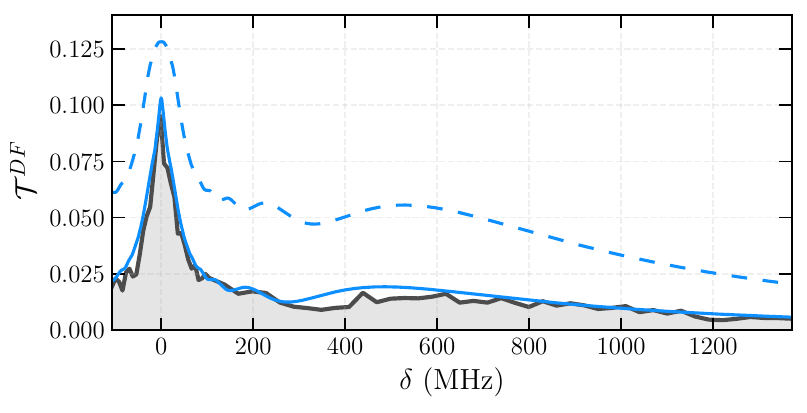}
\caption{Calibration of the spectral model. Solid black line: Spectrum of a partially condensed cloud at $T=\SI{125}{\nano\kelvin}$, with $N_{\mathrm{th}}=1.3\cdot10^5$, and $N_0=4.9\cdot10^5$ obtained from absorption imaging. Dashed blue line: Model spectrum assuming infinite aperture radius and $\alpha=1$. Solid blue line: Model spectrum with the calibrated values for NA and $\alpha$.}
\label{fig. Calibration}
\end{figure}
The result of the calibration is shown in~\cref{fig. Calibration}, where one calibration spectrum is compared with two theoretical spectra; one with $\alpha=1$ and infinite aperture radius, and another with the optimized values. The same calibration result is used for BF spectra. The cloud parameters used in the model were obtained from the TOF images and are not fitted spectroscopically. The calibration results in significantly improved agreement between the expected and measured spectra.

\section{Spectroscopic detection}
\label{sec. Spectroscopic fitting}

This section presents the spectra and benchmarks the extracted cloud parameters – $N_0$, $N_{\mathrm{th}}$, and $T$ – against time-of-flight imaging data. Each spectrum and the corresponding TOF image are taken on the same cloud, and the experiment is repeated at different points in the evaporation sequence to probe various regimes, ranging from thermal clouds to condensate fractions of $97\%$. The light intensity is chosen such that a maximum of 10\% of atoms are lost in the most shallow trap and the coldest clouds during the measurement of a spectrum, while the destructivity is smaller for all other measurements. The heating is limited by the trap depth since all measurements are performed at a trap depth corresponding to the final depth of evaporative cooling. 

\subsection{Observed spectra}
\label{subsec. Illustrations of spectroscopic detection}

\begin{figure*}[t]
\centering
\includegraphics[width=\linewidth]{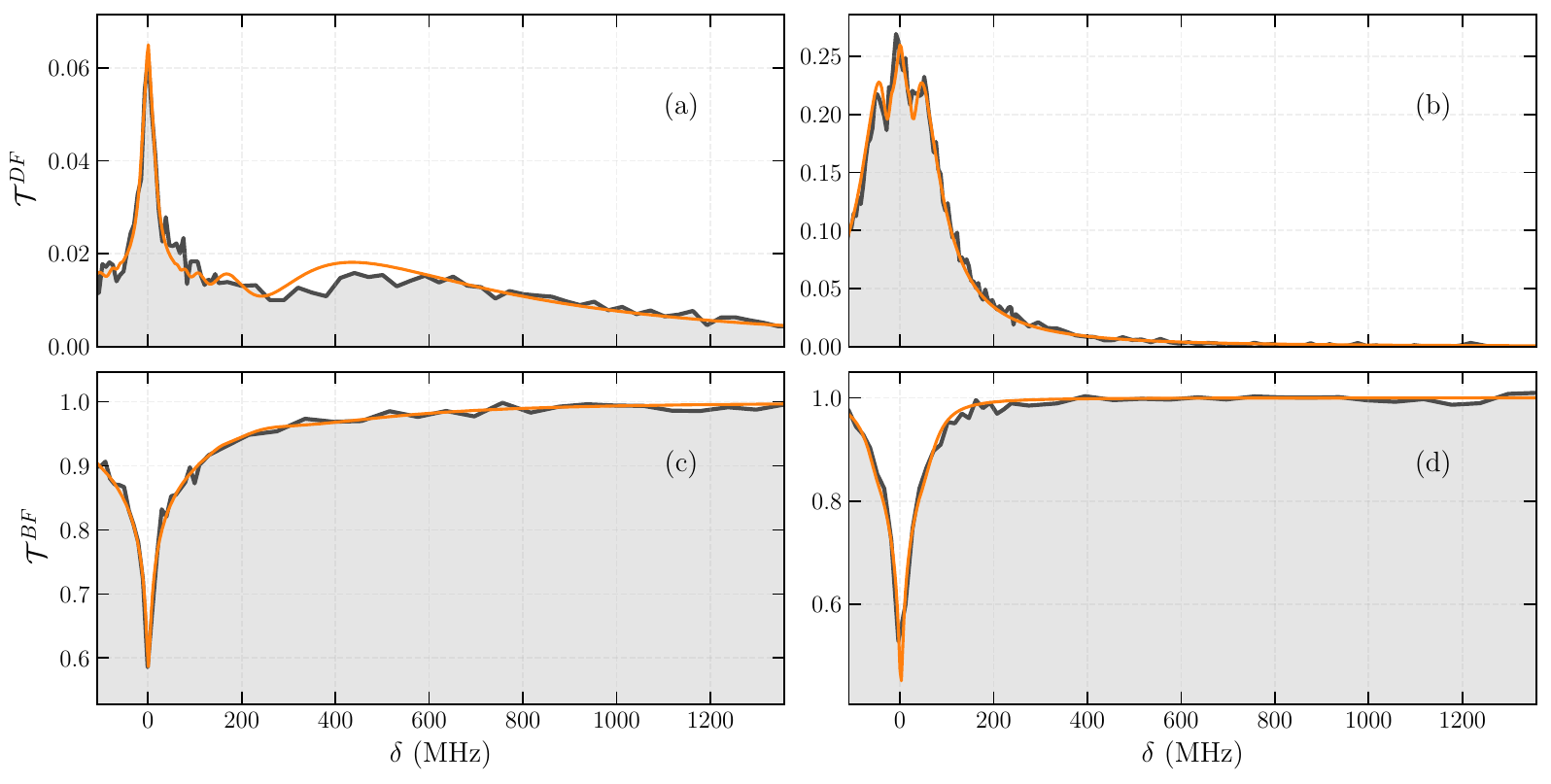}
\caption{Spectroscopic fit examples. (top) Dark-field spectra for a partially condensed cloud (a) and a small, purely thermal cloud (b). The bimodal dark-field spectrum shows a narrow thermal peak on top of a broad non-monotonous background caused by the BEC. (bottom) Bright field spectra for a partially condensed cloud (c) and a small, purely thermal cloud (d). The bright-field spectra both display monotonous signals. However, the partially condensed signal has a broader absorption background.}
\label{fig. exemplary fits}
\end{figure*}
\Cref{fig. exemplary fits} shows examples of two DF and two BF measurements: a cold, bimodal cloud and a warm, purely thermal cloud. The figure shows the normalized transmission spectra according to~\cref{eq. DF raw data to normalized T} and~\cref{eq. BF raw data to normalized T} for DF and BF, respectively.

The DF spectrum in~\cref{fig. exemplary fits}(a) displays the discussed features with characteristic local minima and maxima. Most noticeable is the broad peak in signal at $\delta\approx\SI{500}{\mega\hertz}$, which is the point where the phase-shift causes maximal signal, also discussed in \cref{subsec. Spectral theory}. The warmer, purely thermal cloud in~\cref{fig. exemplary fits}(b) displays a similar feature at $\delta\approx\SI{50}{MHz}$. This always occurs if the phase-shift exceeds $\pi$ at any point in the spectrum, and the detuning $\delta$ of this peak grows with increasing optical density.

The BF spectra in~\cref{fig. exemplary fits}(c)-(d) do not display such local minima, since the method is insensitive to the phase-shift of the probe light. By comparing the cold BF signal to the warmer, purely thermal BF signal, it is nonetheless clear that the dense BEC contributes a broad, shallow absorption signal.

The DF spectra are fitted with~\cref{eq. normalized T DF} multiplied by $\mathcal{P}^{DF}$ and the BF spectra with~\cref{eq. normalized T BF} multiplied by $\mathcal{P}^{BF}$ where $\mu$, $T$ and the spectral resonance frequency are free fitting parameters. These fits are shown in~\cref{fig. exemplary fits}, along with the experimental spectra.

The structures in the cold DF spectra are well represented by the theoretical model. Specifically, the narrow peak at resonance caused by the thermal component, as well as the broad spectral feature caused by the condensed component, is captured by the fit. Moreover, the position of the local maxima in the spectra is captured by the fitted model.

However, there remains a small systematic deviation between the model and the experimental data around the largest local maxima at $\delta\approx\SI{500}{\mega\hertz}$. This is likely caused by the imperfections in the model, such as the use of the thin-lens approximation and the validity of the effective aperture model. In addition, further effects beyond single-atom model, such as induced dipole-dipole interactions and bosonic enhancement, may contribute to the signal as discussed in~\cref{sec. Conclusion and outlook}.

Overall, the fitted spectra replicate the experimental spectra well at both temperatures and thus illustrate the robustness of the model in both the DF and the BF configuration.

Note that clouds with peak $od$ much less than one cannot be characterized in this way, since the signal reduces to a linear regime and reveals no spatial information. In the trap considered here, this regime would be reached around 1000 atoms, and it would require fewer atoms in tighter traps.

\subsection{Comparison with time-of-flight imaging}
\label{subsec. Comparison to time-of-flight imaging}
\begin{figure*}[t]
\centering
\includegraphics[width=\linewidth]{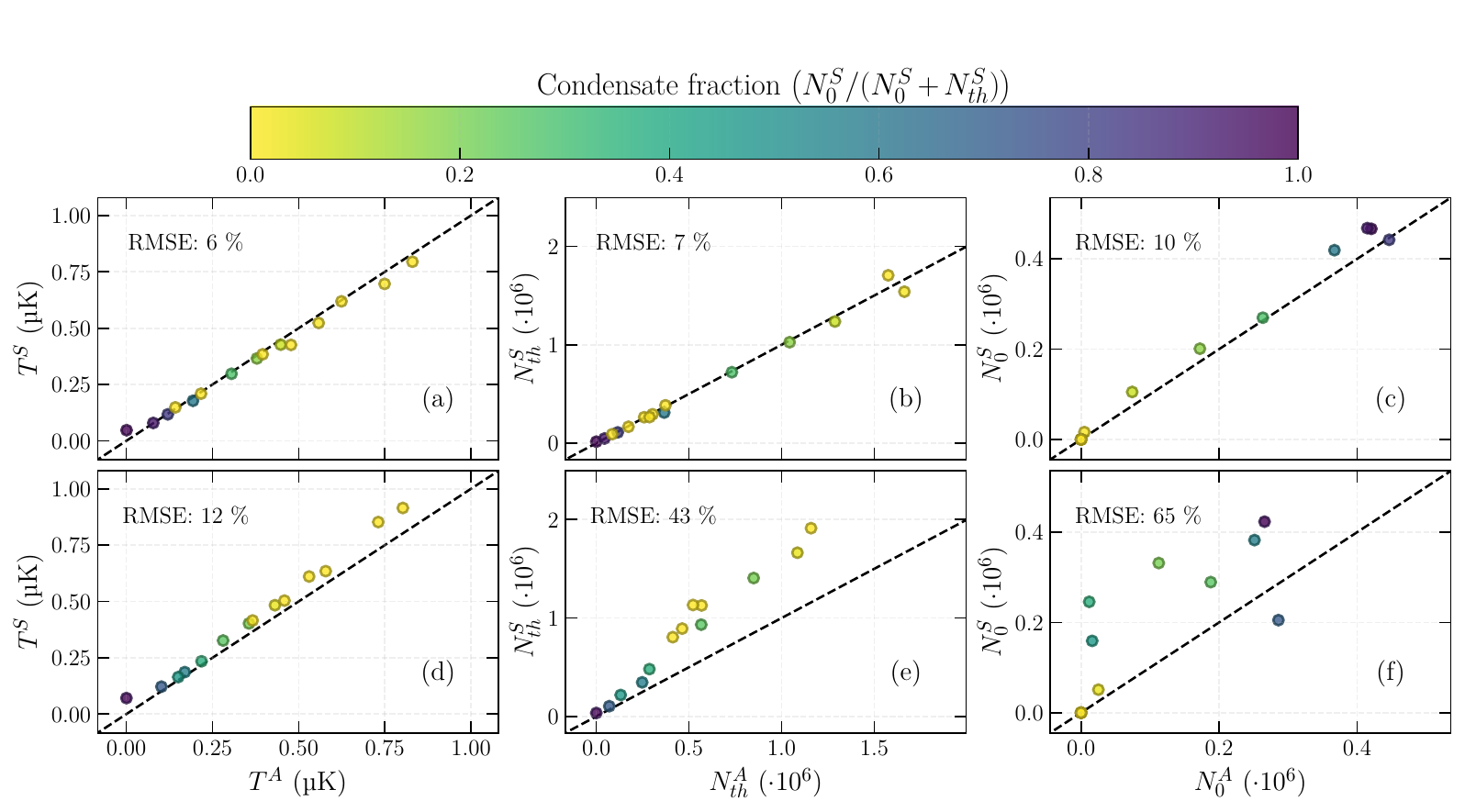}
\caption{Comparison between absorption imaging and spectroscopy. (a)-(c) DF results and (d)-(e) BF results compared to the temperature, thermal atom number in the second column, and condensed atom number obtained from absorption imaging. The color scale indicates the condensed fraction as calculated from the spectral fits. The overall agreement in each panel is indicated with the root-mean squared error (RMSE). For $N_0$ only condensate fractions above $\SI{0.1}{}$ are considered. Fits of the coldest cloud are illustrated in~\cref{fig. imaging-vs-spectrum}.}
\label{fig. errorplot}
\end{figure*}
To investigate the accuracy of spectral fits, the method is compared to absorption images after TOF. A large parameter range of atom numbers, temperatures, trap frequencies, and condensate fractions is used to compare the parameters obtained from the two methods. In the following, parameters extracted from spectra are indicated with a superscript $S$, and parameters extracted from absorption images are indicated with a superscript $A$. The differences between the two methods are given by
\begin{align}
\nonumber\Delta T &= T^S-T^A,\\
\nonumber\Delta N_{th}&=N_{th}^S-N_{th}^A,\\ \Delta N_0&=N_0^S-N_0^A.
\end{align}

\Cref{fig. errorplot} shows the comparison between the two methods. Consider first \cref{fig. errorplot}(a), which shows the temperature based on the DF method. The DF results agree with absorption imaging within $\SI{6}{\percent}$ RMSE for all points except the coldest with the highest condensed fraction. The disagreement here is not caused by the spectroscopic fits but instead by the absorption imaging method, which is unable to detect the small remaining thermal component (see \cref{subsec. Advantages of spectroscopic detection}). A similar result is obtained for the comparison of thermal atom number estimation, as shown in \cref{fig. errorplot}(b). Again, the two methods agree within $\SI{7}{\percent}$ for all points except for the coldest cloud. Finally, the comparison of the condensed atom number for this case is shown in~\cref{fig. errorplot}(c), which displays agreement within $\SI{10}{\percent}$ for condensate fractions above $\SI{0.1}{}$. This agreement validates the applicability of the DF method to extract cloud parameters over the entire parameter range. The observed discrepancies fall within the typical systematic uncertainties of both techniques~\cite{Vibel2024}.

\Cref{fig. errorplot}(d)-(f) show a comparison of the three parameters $T$, $N_{th}$, and $N_0$ between the BF method and the TOF detection. In this case, the temperature estimates deviate on average by $\SI{12}{\percent}$. The estimation of thermal atom number using the BF method is less accurate, especially for larger clouds, resulting in an average deviation of $\SI{43}{\percent}$. Finally, BF spectra perform poorly at estimating the BEC atom number with deviations of $\SI{65}{\percent}$ on average for condensed fractions above $\SI{0.1}{}$. This is mainly caused by the limited numerical aperture in combination with the fact that BF is not sensitive to the phase-shift. Thus, the BF signal caused by the BEC component is largely lost. Note, however, that even the BF method is able to measure the presence of a thermal component at the lowest temperature, even when the TOF detection fails to do so.

In summary, the DF method shows good agreement with the TOF imaging method for estimation of all three parameters, and even displays increased sensitivity to the small thermal component at large  BEC fraction compared to TOF imaging. The method works with an imaging system that is not designed to resolve in-situ density distributions. The BF method works equally well for estimating temperature, but shows increased noise and less accuracy for atom number estimation.

\subsection{Low temperature spectroscopic detection}
\label{subsec. Advantages of spectroscopic detection}

\begin{figure}[t]
\centering
\includegraphics[width=\linewidth]{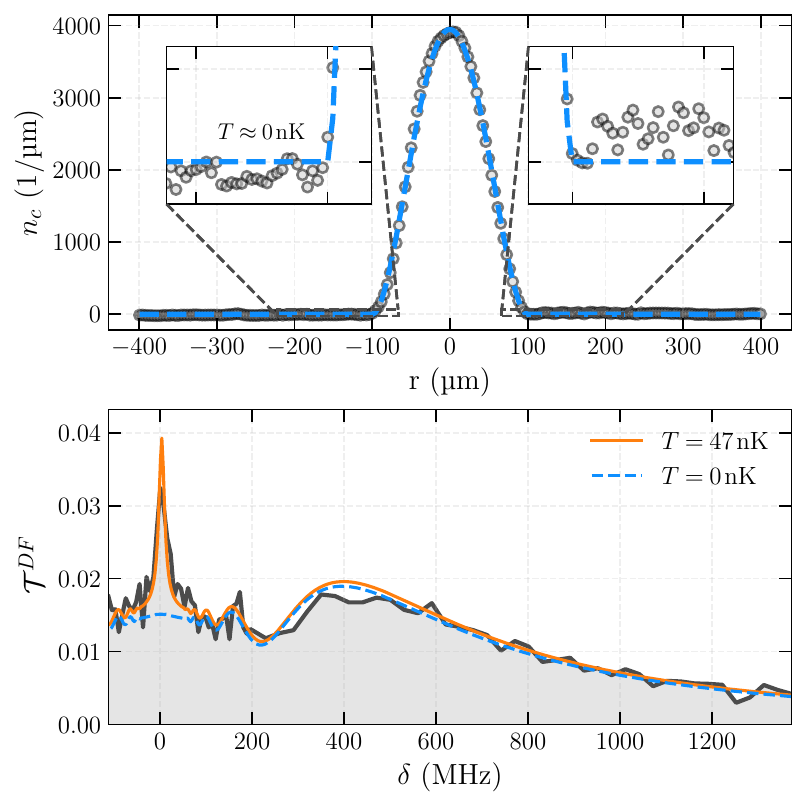}
\caption{Comparison between absorption imaging after $\SI{36}{ms}$ time of flight and a DF spectrum for a very cold cloud. (top) The integrated atom number density shows no sign of a thermal component. The blue dashed line is a fit of the BEC component according to the Thomas-Fermi model. (bottom) The spectroscopic method shows a clear signal due to the thermal component at resonance and a broad feature due to the BEC. The solid orange line is a fit with the spectroscopic model. The blue dashed line is the spectrum of the BEC according to the parameters extracted from the top panel. Data were obtained from averages of 18 realizations.}
\label{fig. imaging-vs-spectrum}
\end{figure}

Spectroscopy thus provides an alternative characterization method for Bose-Einstein condensates at very low temperatures with a small thermal fraction. This is demonstrated in~\cref{fig. imaging-vs-spectrum} by comparing absorption images and DF measurements of the coldest cloud in our measurements. The top panel shows the integrated atom number density along two dimensions for 18 averaged images, and the bottom panel shows the average spectroscopic signals of the same clouds.

The absorption images do not contain a thermal tail even after averaging 18 images, as shown in \cref{fig. imaging-vs-spectrum} and thus a bimodal fit is not possible. Two insets in the panel enlarge the region around the tail of the cloud, demonstrating the lack of thermal signal. Thus, it is not feasible to extract the thermal cloud parameters, and the cloud is only fit with a Thomas-Fermi distribution.

On the contrary, the spectrum shows a clear signal due to both components, and a fit yields a temperature of $T=\SI{47}{\nano\kelvin}$, corresponding to a condensed fraction of $\SI{97}{\percent}$. To illustrate which part of the signal arises from the small thermal component of the cloud, a calculated spectrum with the appropriate BEC atom number but no thermal cloud is also included. Note that the small number of thermal atoms provides a clear signal close to resonance in the spectrum, which is well distinguished from the BEC signal.

\begin{figure}[t]
\centering
\includegraphics[width=\linewidth]{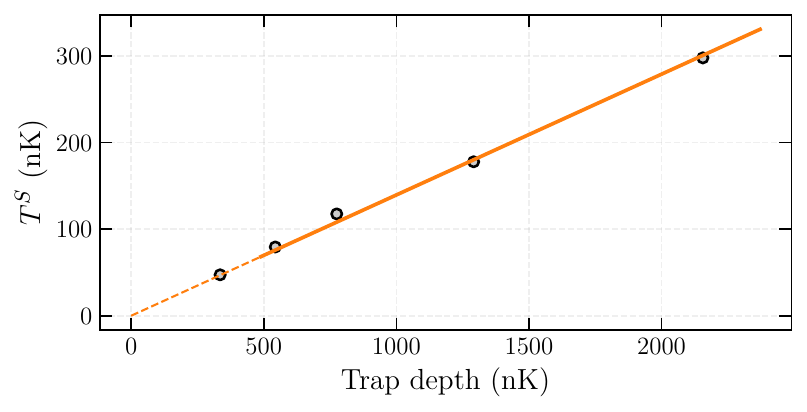}
\caption{Spectroscopically measured temperature as a function of dipole trap depth. The solid line displays a fit to the data excluding the coldest sample. The dashed line is an extrapolation of the fit.}
\label{fig. confirm low T}
\end{figure}
To verify this low temperature measurement \cref{fig. confirm low T} shows a comparison between the spectroscopically measured temperature and the dipole trap depth. This temperature measurement corresponds to the one shown in \cref{fig. errorplot}(a), which is in good agreement with temperatures obtained from absorption images. The observed temperature is fitted with a linear relationship, excluding the coldest measurement. An extrapolation of this fit is in good agreement with the measurement at low temperature, thereby validating the spectroscopic measurement in the regime where absorption imaging fails.

Whilst measurements using absorption imaging for high condensate fractions have been reported~\cite{Leanhardt2003}, resolving the thermal signal remains challenging due to the vanishing non-condensed fraction. This can be mitigated by increasing the time-of-flight duration before imaging, allowing the condensed and thermal components to separate more clearly. Such long expansion times are, for example, used in low-gravity experiments, where expansion times of up to $\SI{400}{\milli\second}$ are realized~\cite{temperatureLowGravity}. This is not easily feasible in typical experiments, while the spectroscopic method proposed here can be integrated into most experimental setups.

\section{Conclusion and outlook}
\label{sec. Conclusion and outlook}

In conclusion, we have performed spectroscopic measurements of partially condensed BECs in a dark-field configuration, revealing rich spectral structures. A theoretical description based on a simple linear response model was developed, which explains the observed structures and allows for the extraction of cloud parameters. We also performed spectroscopic measurements in a bright-field configuration and demonstrated the superiority of the dark-field technique. 

Based on the model, a complementary spectroscopic detection technique was demonstrated, which enables in-situ minimally destructive measurements of the cloud parameters without requiring spatially resolved imaging. The method relies on few-photon detection with a frequency-agile laser system, which allows the probe light frequency to be scanned across a broad frequency range. 

The parameters extracted spectroscopically were benchmarked against time-of-flight absorption imaging across a range of experimental parameters. In particular, the dark-field method provides reliable estimates of condensed and thermal atom numbers as well as temperature across a large parameter range. On average, spectroscopic measurements and absorption imaging agree within $\pm6\%$ for temperature, $\pm7\%$ for thermal atom number, and $\pm10\%$ for condensed atom number. 

Importantly, the spectral approach demonstrated enhanced sensitivity to the small thermal components in clouds with high condensate fractions, where typical absorption imaging failed to extract the temperature. Since the spectra can be measured in-situ without the need for high-resolution imaging optics, the method also avoids the need to model the expansion dynamics. For those reasons it may be of benefit to research with ultracold atoms in general.

In future work, the model introduced here can be improved to achieve higher precision and avoid the need for calibration with absorption imaging. In particular, the thin-lens approximation was used to estimate the effect of the atoms on the light field, thus avoiding a detailed simulation of light propagation through the cloud. The agreement between the model could potentially be improved by implementing a more detailed method, such as the one presented in~\cite{Deb_2020}.

Furthermore, the spectroscopic measurements are, in principle, susceptible to many-body atom-light interaction effects. This includes effects, such as bosonic enhancement~\cite{Konstantinou_2026, morice1995refractive} and dipole-dipole interaction~\cite{spectrumPaper2_Browaeys2014, spectrumPaper1_Browaeys2016, DipoleDipoleCloudShapeChange}. In fact, dipole-dipole interactions are expected to play a role in these measurements since they become important at densities $\rho>1/\lambda^{3}$, where $\lambda$ is the wavelength of the probe light, and the peak density of atoms in these measurements is greater than $100/\lambda^{3}$. So far, these effects have not been observed in our data. Their inclusion in our model would therefore be of interest to evaluate their effect.

\section*{Acknowledgments}
\label{sec. Acknowledgments}
We acknowledge support from the Danish National Research Foundation through the Center of Excellence “CCQ” (DNRF152) and by the Novo Nordisk Foundation NERD grant (Grantno. NNF22OC0075986). N.K. was supported by the Marsden Fund (Contract No. UOO2421).

\bibliography{bib}

\end{document}